\documentclass[aps,prb,11pt,tightenlines,superscriptaddress,preprint]{revtex4}
\usepackage{graphicx}
\newcommand{\supercell}[1]{#1$\times$#1$\times$#1 supercell} 
\newcommand{\supcell}[1]{#1$\times$#1$\times$#1} 

\begin{document}
\title{Bulk and surface energetics of lithium hydride crystal: \\
benchmarks from quantum Monte Carlo and quantum chemistry}
\author{S. J. Binnie}
\affiliation{Thomas Young Centre, UCL,
  London WC1E 6BT, UK} 
\affiliation{Dept. of Physics and Astronomy, UCL,
  London WC1E 6BT, UK} 
\affiliation{London Centre for Nanotechnology, UCL,
  London WC1H 0AH, UK} 

\author{S. J. Nolan}
\affiliation{Centre for Computational Chemistry, School of Chemistry,
  University of Bristol, Bristol BS8 1TS, UK} 
\author{N. D. Drummond}
\affiliation{Theory of Condensed Matter Group, Cavendish Laboratory,
  University of Cambridge, Cambridge CB3 0HE,
  UK}

\author{D.~Alf\`{e}}
\affiliation{Thomas Young Centre, UCL,
  London WC1E 6BT, UK} 
\affiliation{Dept. of Physics and Astronomy, UCL,
  London WC1E 6BT, UK} 
\affiliation{London Centre for Nanotechnology, UCL,
  London WC1H 0AH, UK} 
\affiliation{Dept. of Earth Sciences, UCL,
  London WC1E 6BT, UK} 

\author{N. L. Allan}
\author{F. R. Manby} 
\affiliation{Centre for Computational Chemistry, School of Chemistry,
  University of Bristol, Bristol BS8 1TS, UK}  

\author{M. J. Gillan}
\affiliation{Thomas Young Centre, UCL,
  London WC1E 6BT, UK} 
\affiliation{Dept. of Physics and Astronomy, UCL,
  London WC1E 6BT, UK} 
\affiliation{London Centre for Nanotechnology, UCL,
  London WC1H 0AH, UK} 

\date{\today}


\begin{abstract}
We show how accurate benchmark
values of the surface formation energy of crystalline lithium hydride can be computed by
the complementary techniques of quantum Monte Carlo (QMC) and
wavefunction-based molecular quantum chemistry. To demonstrate
the high accuracy of the QMC techniques, we present a detailed study
of the energetics of the bulk LiH crystal, using both pseudopotential
and all-electron approaches. We show that the equilibrium lattice parameter
agrees with experiment to within $0.03~$\%, which is around the experimental uncertainty, and the cohesive energy
agrees to within around $10$~meV per formula unit. QMC in periodic slab
geometry is used to compute the formation energy of the 
LiH~(001) surface, and we show that
the value can be accurately converged with respect to slab thickness
and other technical parameters.
The quantum chemistry calculations build on the recently 
developed hierarchical scheme
for computing the correlation energy of a crystal to high precision.
We show that the hierarchical scheme allows the accurate calculation
of the surface formation energy, and we present results that are well converged with respect
to basis set and with respect to the level of correlation treatment. The 
QMC and hierarchical results for the surface formation energy agree to
within about $1$~\%.
\end{abstract}

\maketitle

\section{Introduction}

\label{sec:intro}
The surface formation energies of materials are key quantities in fields
as diverse as nanotechnology, mineral science and fracture mechanics. However,
the accurate measurement of surface energies is fraught with
difficulties, so there is often a need to rely on 
calculated values. In principle,
electronic-structure methods based on density-functional theory (DFT)
should be capable of giving reliable surface energies, but in practice
it is found that computed values depend strongly on the approximation
used for the exchange-correlation 
energy.~\cite{goniakowski96,mattsson_computing_2002,yu_first-principles_2004,binnie_benchmarking_2009} 
There are two main kinds of 
electronic-structure technique that allow one to go beyond DFT and
achieve better accuracy: quantum Monte Carlo (QMC), and the
wavefunction-based correlation techniques usually associated with
molecular quantum chemistry (QC). We show here how these two very
different approaches can be used in a complementary way to produce
accurate benchmark values for surface formation energy, using as a
test case the (001) surface of crystalline LiH.

There have been many DFT calculations of the surface formation
energies $\sigma$ of different kinds of materials, including ionic compounds,
covalent semiconductors and metals. In some cases, the variation
of predicted $\sigma$ values with the assumed exchange-correlation
functional has been studied, and it is found that generalized
gradient approximations (GGA) such as PBE and PW91 often give
$\sigma$ values that are $\sim 30$~\% lower than those predicted by
the local density approximation
(LDA).\cite{binnie_benchmarking_2009,alf_energetics_2006,yu_first-principles_2004,mattsson_computing_2002}
Since GGAs are generally 
more accurate than LDA for bonding energies, and since the energy
needed to form a surface would seem to be closely related to the energy
needed to break bonds, it might be expected that GGA values of $\sigma$
would be more accurate. However, in the few cases where there are reliable
experimental data, this expectation is not fulfilled, and the rather
scattered evidence suggests that the LDA may be more accurate.\cite{alf_energetics_2006,yu_first-principles_2004,mattsson_computing_2002} A connection
has been made with the superiority of LDA over GGAs for the surface
energy of jellium.\cite{almeida_surface_2002}

In this rather confused situation, it is helpful to seek ways of computing
benchmark values of $\sigma$ which do not suffer from the uncertainties
of DFT. Quantum Monte Carlo, and specifically diffusion Monte Carlo 
(DMC)~\cite{foulkes_quantum_2001,needs_continuum_2010} 
offers one way of achieving this.
It is well established that DMC is
usually much more accurate than DFT for the energetics of extended
systems, and there are ways of systematically improving its accuracy.
Nevertheless, it is subject to errors that are not completely
controllable, and this is where the methods of molecular
quantum chemistry can play an important role. The electron-correlation
techniques that we use here are mainly second-order M{\o}ller-Plesset theory (MP2)
and the coupled-cluster scheme CCSD(T) (including single, double and a perturbative treatment of
triple excitations). Efforts to apply these QC techniques to the energetics of
extended systems go back many years, particularly using the so-called
incremental approach~\cite{stoll92,paulus_method_2006}. 
More recently, the MP2 approximation has been
implemented for periodic systems in several codes~\cite{kresse_mp2,pisani08}.
The present authors have reported a technique referred to as
the hierarchical method for applying molecular QC methods to
perfect crystals, and for the case of LiH have shown that it can
deliver the cohesive energy to an absolute accuracy of $\sim 5$~meV
per formula unit and the equilibrium lattice parameter to better than
$0.1$~\%.\cite{nolan_calculation_2009,manby_extension_2006}\\

We have chosen to study the surface energetics of LiH, partly because
it is a material for which we expect DMC to give very high accuracy,
and partly because we already know that hierarchical QC is very 
accurate.~\cite{nolan_calculation_2009} The crystal has the rock--salt
structure, and the simplicity of this structure facilitates the calculations.
We have several main aims. First, we want to show that DMC does indeed
deliver high accuracy for the properties of the LiH crystal, particularly
if we use all-electron rather than pseudopotential DMC (pp-DMC). Second, we report
our periodic slab calculations of $\sigma$ for the LiH~(001) surface,
using both pseudopotential and all-electron DMC, and we show
that we can achieve a high degree of convergence with respect to
slab thickness and other technical parameters. Third, we show that the
hierarchical QC scheme that gives such good accuracy for bulk LiH
also provides a practical way of obtaining benchmark values of $\sigma$.
The hierarchical methods allow us to calculate explicitly the contribution
of core-valence correlation to $\sigma$, and we shall see that this is
significant. Naturally, close agreement between the
$\sigma$ values computed by the QMC and QC approaches supports 
the credibility of both, and this will be carefully assessed.\\


\section{Techniques}
\label{sec:techniques}

\subsection{Quantum Monte Carlo}
\label{sec:qmc}

For present purposes, the name quantum Monte Carlo refers to
two techniques for determining the ground-state energy of
a many-electron system (for reviews, see e.g. 
Refs.~\cite{foulkes_quantum_2001,needs_continuum_2010}). 
Our high-precision results are obtained
using diffusion Monte Carlo (DMC), a technique that projects
out the ground state by evolving the many-electron wavefunction
in imaginary time with the aid of an approximate trial wavefunction.
An optimized form of this trial function is computed using
variational Monte Carlo (VMC), which is an implementation of
the variational principle of quantum mechanics. The VMC and DMC
calculations in this work are performed using the {\sc CASINO}
package~\cite{needs_continuum_2010}.

The trial wavefunctions used here have the standard single-determinant
Slater-Jastrow form:
\begin{equation}
\Psi_{\rm T} = D_\uparrow D_\downarrow e^J \; ,
\label{eq:trial_wfn}
\end{equation}
where $D_\uparrow$ and $D_\downarrow$ are up- and down-spin
Slater determinants of single-electron orbitals $\psi_n$. Electron
correlation is approximately described by $J$, which is a sum of
three types of terms: electron-electron terms $u$,
electron-nucleus terms $\chi$, and electron-electron-nucleus
terms $f$. These three terms contain parameters that are optimized
using VMC, so as to make $\Psi_{\rm T}$ as close as possible to
the true ground-state wavefunction. The optimization works
with the local energy 
$E_{\rm L} \equiv \Psi_{\rm T}^{-1} \hat{H} \Psi_{\rm T}$,
where $\hat{H}$ is the many-electron Hamiltonian. We follow
the common procedure of varying the parameters so as to minimize
the variance of $E_{\rm L}$ (the variance would be zero if
$\Psi_{\rm T}$ were the exact ground-state wavefunction). VMC can
be used equally well as an all-electron technique or with non-local
pseudopotentials to represent the interaction between valence electrons
and atomic cores.

The idea of 
DMC~\cite{ceperley_ground_1980,foulkes_quantum_2001,needs_continuum_2010} 
is to represent the exact many-electron wavefunction $\Phi$
as a density of brown\-ian particles, or `walkers'. In the evolution
of the wavefunction according to the time-dependent Schr\"{o}dinger
equation in imaginary time, the optimized approximation
$\Psi_{\rm T}$ from VMC is used to guide the walkers, in a manner
related to importance sampling. DMC aims to stochastically simulate
the diffusion, birth, death and drift of the walkers, which, after
an equilibration period, samples the exact ground-state wavefunction.
In practice, the fermionic nature of electrons prevents DMC from
being completely exact, and the nodal surfaces of the wavefunction
are 
constrained to be those of $\Psi_{\rm T}$ -- this is
the well-known fixed-node 
approximation~\cite{anderson75}. We shall see that this
approximation incurs only small errors in the present work. 
A number of other technical issues have to be addressed,
including time-step errors, pseudopotential errors, the choice and
representation of the single-electron orbitals $\psi_n$, 
and the stability of walker populations, and we 
summarize these next. The treatment of system size errors will
be discussed in Sec.~\ref{sec:bulk_and_surface}.

The walkers propagate by using the approximate small-time-step 
Green's function as a transition probability in configuration 
space. The approximate Green's function also includes a term 
that gives a probability for a given walker to `branch' (become
two walkers) or to be discarded entirely.
The use of a discrete time-step incurs errors,
but these can be rendered negligible by the usual procedure of
extrapolating to the zero-time-step limit. We shall present 
both pseudopotential and all-electron DMC calculations
on the LiH bulk and surface. For the pseudopotential work,
we use the Dirac-Fock pseudopotentials due to 
Trail and Needs~\cite{trail_smooth_2005,trail_norm-conserving_2005}.
It is difficult to treat non-local
pseudopotentials in DMC, and we employ the usual locality 
approximation~\cite{mitas_nonlocal_1991},
which introduces errors proportional to the square of the difference
between $\Psi_{\rm T}$ and the exact ground-state wavefunction $\Phi$. The
comparison of our pseudopotential and all-electron results will
help us to quantify these errors.

The single-electron orbitals $\psi_n$ used in the trial wavefunction
$\Psi_{\rm T}$ (see Eq.~(\ref{eq:trial_wfn})) were 
generated by DFT calculations
with the LDA functional. We make this choice because there is considerable
evidence~\cite{kent1998,ma2009} that this gives 
a $\Psi_{\rm T}$ that is closer to the
true ground state. The $\psi_n$ were computed by plane-wave
calculations with the {\sc quantum espresso} 
package~\cite{QE-2009}. However, the
direct use of $\psi_n$ in a plane-wave representation in DMC
is very inefficient, and instead we re-expand the $\psi_n$ in a 
blip-function (B-spline) basis~\cite{alf_efficient_2004}, 
using the standard relation between
the blip-grid spacing and the plane-wave cut-off. In the case of all-electron
DMC, a further modification is necessary, since it is crucially
important that $\Psi_{\rm T}$ has the correct electron-nuclear
cusp at the nuclear positions. The technique we have used to ensure
this with the blip basis is described in Appendix~A.

Since walkers can branch or be discarded after each step, the 
walker population fluctuates. A reference energy in the 
approximate Green's function 
allows us to bias the branching, and thus control the 
population. However, in regions of particularly 
low energy (especially divergences at point charges), this 
mechanism is not enough, and a walker trapped in this 
region (and its offspring) can branch repeatedly, causing 
a population explosion which destroys the statistics of subsequent moves.\\

\subsection{QMC for bulk and surface energies}
\label{sec:bulk_and_surface}

Correction for errors due to the limited size of the periodically
repeated cell is important in the calculation of both
bulk and surface energies. As usual, we distinguish between
single-particle and many-body errors. The former are due to the
fact that $k$-point sampling cannot be performed with DMC, and are
analogous to those that would arise in single-particle methods such
as DFT without $k$-point samplng; the latter are due to
the spurious interaction of electrons with their periodic images.
To correct for single-particle errors, we use the 
formula~\cite{foulkes_quantum_2001}:
\begin{equation}
E_\infty = E_{\rm cell}^{\rm DMC} +
a ( E_\infty^{\rm DFT} - E_{\rm cell}^{\rm DFT} ) \; ,
\label{eq:single}
\end{equation}
where $E_{\rm cell}^{\rm DMC}$ and $E_{\rm cell}^{\rm DFT}$ are the
energies of the given cell with DMC and DFT (no $k$-point sampling
with DFT), $E_\infty^{\rm DFT}$ is the DFT energy of the cell with
perfect $k$-point sampling, and $E_\infty$ is the corrected DMC energy;
if enough data for different system sizes are available, $a$ can
be treated as a fitting parameter.

One way of correcting for many-body size errors is to use a modified
form of the Coulomb interaction known as the Model Periodic Coulomb
(MPC) interaction in the DMC calculations.~\cite{fraser_finite-size_1996,williamson_elimination_1997,kent_finite-size_1999} We used this technique,
in combination with Eq.~(\ref{eq:single}) for our all-electron calculations
on bulk LiH. An alternative approach is the scheme due to Kwee
{\em et al.}~\cite{kwee_finite-size_2008}, which corrects for both single--particle and many--body
errors in a single formula:
\begin{equation}
\label{eq:kwee}
E_\infty=E_N^{\rm DMC}+E_\infty^{\rm LDA}-E_N^{\rm KZK} \; ,
\end{equation}
which somewhat resembles Eq.~(\ref{eq:single}). Here,
$E_{\rm cell}^{\rm KZK}$ is a DFT-like energy of the cell (no $k$-point
sampling), which uses a functional designed to mimic the sum of
single-particle and many-body errors, while $E_\infty^{\rm LDA}$ is the
same as $E_\infty^{\rm DFT}$ in Eq.~(\ref{eq:single}), evaluated with
the LDA functional. We used this scheme of Kwee {\em et al.} for the
pseudopotential calculations on the bulk. Whichever method is used
to correct for the many-body size errors, in the case of the bulk
calculations we apply a further two-point extrapolation to
remove residual finite-size errors. This extrapolation employs
the formula:
\begin{equation}
\label{eq:twopnt}
 E_\infty = \left( N E_N - M E_M \right) \left/ \left( N - M \right) \right.
\; ,  
\end{equation}
where $E_N$ and $E_M$ are the DMC energies per formula unit of 
supercells containing $N$ and $M$ formula units 
respectively.\cite{ceperley_ground_1980,ceperley_ground_1987,rajagopal_variational_1995,zong_spin_2002}

Our DMC calculations of the surface formation energy are performed in
slab geometry, so that we work with slabs having infinite
extent in the plane of the surface and having 
a specified number $N$ of ionic layers. Periodic boundary conditions
are applied in the surface plane, so that we have supercell geometry
only in two dimensions. With the blip basis set used for the present
work, it is unnecessary to apply periodic boundary conditions in
the direction perpendicular to the surface.

As usual, the surface formation energy $\sigma$ is the work needed to
create an area $A$ of new surface, starting from the perfect bulk 
crystal, divided by $A$. In slab geometry, if $E_{\rm slab} ( N )$
is the energy per supercell of the $N$-layer slab, 
$\nu_{\rm slab} ( N )$ is the number of formula units per supercell
of the $N$-layer slab, and $e_{\rm bulk}$ is the energy per formula
unit of the bulk crystal, then $\sigma$ is given by:
\begin{equation}
\sigma = \lim \left( E_{\rm slab} ( N ) - 
\nu_{\rm slab} ( N ) e_{\rm bulk} \right) / A \; ,
\label{eq:simpsurf}
\end{equation}
where $A$ is the total surface area (both faces) per supercell of the slab.
In Eq.~(\ref{eq:simpsurf}), we must take the limit as the 
number of layers $N$ and the
surface dimensions of the supercell both tend to infinity. For comparison with
experimental data, the ionic positions in the slab should also 
be relaxed to equilibrium, but in the present work we are concerned
mainly with comparing different theoretical approaches, and we focus
on the unrelaxed value of $\sigma$, for which all ions in the slab
have their bulk positions.

Instead of using Eq.~(\ref{eq:simpsurf}) directly, we prefer to 
use the well-known procedure of extracting $\sigma$ from a series
of slab calculations of increasing $N$, using the fact
that as $N \rightarrow \infty$, $E_{\rm slab}$ has the asymptotic
form:
\begin{equation}
E_{\rm slab} ( N ) \rightarrow A \sigma + N E_{\rm layer} \; .
\label{eq:surf}
\end{equation}
Here, $\sigma$ is the surface formation energy with the chosen
surface supercell, and $E_{\rm layer}$ is the bulk energy per ionic
layer with this supercell. Eq.~(\ref{eq:surf}) is 
equivalent to Eq.~(\ref{eq:simpsurf}), but the
extraction of $\sigma$ for a given surface supercell 
from Eq.~(\ref{eq:surf}) is usually
more robust. We note that the value of $E_{\rm layer}$ can be cross--checked
against independent calculations on the bulk crystal, since
$N E_{\rm layer} / \nu_{\rm slab} ( N )$ should be very close to
$e_{\rm bulk}$. 

When correcting for finite--size errors in slab geometry, 
compensation for many--body errors poses technical problems, and
we therefore used only the single--particle
correction of Eq.~(\ref{eq:single}).

\subsection{Correlated quantum chemistry}
\label{sec:techqc}

We show here how the hierarchical 
method~\cite{manby_extension_2006,nolan_calculation_2009}, 
originally developed to treat
bulk crystals, can be used to calculate surface formation energy.
We recall that
the hierarchical method begins by separating
the total energy $e^{\rm tot}$  per primitive cell of a crystal
into Hartree-Fock and correlation parts:
\begin{equation}
e^{\rm tot} = e^{\rm HF} + e^{\rm corr} \; .
\label{eqn:tot_HF_corr}
\end{equation}
The correlation energy $e^{\rm corr}$ is further separated into a molecular contribution and the
so-called ``correlation residual'':
\begin{equation}
e^{\rm corr} = e_{\rm mol}^{\rm corr} + \Delta e^{\rm corr} \; .
\label{eqn:corr_mol_resid}
\end{equation}
In the case of a compound AB having the rock-salt structure, 
$e_{\rm mol}^{\rm corr}$ is the correlation contribution to the binding energy of the AB
molecule, with the bond length taken equal to the nearest-neighbour
distance in the crystal.

The hierarchical method works by combining energies of a sequence of 
finite clusters~\cite{manby_extension_2006,nolan_calculation_2009} 
in such a way as to eliminate surface effects.
For the rock-salt structure, we take cuboidal clusters having
$l$, $m$ and $n$ ions along the three perpendicular edges. By
conventional quantum chemistry techniques, we can compute accurately the
total energy $E_{l m n}^{\rm tot}$ of each $l \times m \times n$ cluster,
which is then decomposed into Hartree-Fock, molecular and residual parts:
\begin{equation}
E_{l m n}^{\rm tot} = 
E_{l m n}^{\rm HF} + \frac{1}{2} l m n \, e_{\rm mol}^{\rm corr} +
\Delta E_{l m n}^{\rm corr} \; .
\label{eqn:totlmn}
\end{equation}
The total energy per primitive cell in the infinite crystal is then:
\begin{equation}
e^{\rm tot} = 
\lim_{l, m, n \rightarrow \infty} \frac{2}{l m n} E_{l m n}^{\rm tot} =
e^{\rm HF} + e_{\rm mol}^{\rm corr} +
\lim_{l, m, n \rightarrow \infty} \frac{2}{l m n} \Delta E_{l m n}^{\rm corr} 
\; .
\label{eqn:tot}
\end{equation}
We calculate the Hartree-Fock contribution $e^{\rm HF}$ using 
standard periodic codes, and $e_{\rm mol}^{\rm corr}$ is
obtained by conventional quantum chemistry techniques. To perform
the limiting process in the third term on the right, the hierarchical method
expresses $\Delta E_{l m n}^{\rm corr}$ as:
\begin{eqnarray}
\Delta E_{l m n}^{\rm corr} & = & 8 E^{000} +
4 [ ( l - 2 ) + ( m - 2 ) + ( n - 2 ) ] E^{001} \nonumber \\
& + & 2 [ ( m - 2 ) ( n - 2 ) + ( n - 2 ) ( l - 2 ) + 
( l - 2 ) ( m - 2 ) ] E^{011} \nonumber \\
& + & ( l - 2 ) ( m - 2 ) ( n - 2 ) E^{111} \; ,
\label{eqn:residlmn}
\end{eqnarray}
where the coefficients $E^{000}$, $E^{001}$, $E^{011}$ and $E^{111}$
represent the energies of corner, edge, face and bulk sites, respectively.
[Note that the definitions of $E^{011}$, $E^{001}$ and $E^{000}$ are
affected by our decision to use factors $(l - 2 ) ( m - 2 ) ( n - 2 )$,
$( m - 2 ) ( n - 2 )$, etc, rather than $l m n$, $m n$, etc. The reason
for making this particular choice of factors is discussed in 
Ref.~\cite{nolan_calculation_2009}.]

Our procedure for obtaining the values of the coefficients in the limit
of infinite $l$, $m$, $n$ requires us to extract $E^{000}$, $E^{001}$,
$E^{011}$ and $E^{111}$ from sets of four independent clusters, and then
systematically to increase the size of the clusters in these sets, as
described in detail in Ref.~\onlinecite{nolan_calculation_2009}. 
For the cohesive energy,
only the limiting value of $E^{111}$ is needed, since 
$\Delta e^{\rm corr} = 2 E^{111}$. However, the procedure also yields the
limiting values of $E^{011}$, $E^{001}$ and $E^{000}$. The value of
$E^{011}$ can be used to obtain the value of the unrelaxed surface formation
energy $\sigma$.

The coefficient $E^{111}$ is the contribution to the energy of a large cluster from an atom in the interior; $E^{011}$ is the same for an atom on the surface. When a surface is formed by opening a gap in the crystal, each atom in the newly formed surface contributes $E^{011}-E^{111}$ to the energy difference. The area of the surface occupied by each atom is $a^2/2$, so the correlation contribution to the formation energy of a new surface is $2( E^{011} - E^{111} ) / a^2$ per ion. We therefore obtain
%
\begin{equation}
\sigma = \sigma^{\rm HF} + 4 ( E^{011} - E^{111} ) / a^2 \; .
\label{eqn:sigma_4}
\end{equation}
We compute the Hartree-Fock part $\sigma^{\rm HF}$ using standard
periodic codes (details will be given later).

\subsection{Zero-point corrections}
Both quantum Monte Carlo and quantum chemistry techniques employ
static calculations, and ignore zero-point energies.
In this work, in order to facilitate
comparison with experiment, all bulk calculations are corrected for
zero-point energy. These corrections are calculated using DFT and the linear
response method. The PBE functional\cite{perdew_generalized_1996} 
is used, since this is known to
give accurate phonon frequencies; our tests with the LDA functional
showed little change in the zero-point energy. The calculations were
performed using the {\sc Quantum Espresso} package~\cite{QE-2009}.\\

\section{Bulk LiH with quantum Monte Carlo}\label{results}
\label{sec:bulk_lih}

We present first our pseudopotential DMC calculations on the bulk,
which already give quite high
accuracy, and also provide valuable 
information about the effect of system-size
errors on the cohesive energy $E_{\rm coh}$ (the energy per
formula unit relative to free atoms), the equilibrium
lattice parameter $a_0$ and the bulk modulus $B$. The all-electron
DMC bulk calculations reported at the end of this Section will show
that an explicit treatment of core-valence correlation improves the
accuracy still further.\\

\subsection{Pseudopotential calculations}
\label{sec:pseudo}

The Dirac-Fock non-local 
pseudopotentials~\cite{trail_smooth_2005,trail_norm-conserving_2005} 
that we use are rather hard,
and we found that a plane-wave cut-off of $4080$~eV and a correspondingly
fine blip-grid spacing was needed to produce accurate orbitals. 
It proved straightforward to eliminate DMC
time-step errors: a time-step of $0.025$~au reduced the error below
$1.0$~meV per formula unit, which is much greater accuracy than we need.

\begin{figure}[h]
  \centering
  \caption{Cohesive energy as a function of primitive
    cell volume for the extrapolated pp-DMC calcualtions and the
    quantum chemistry calculations with and without core-valence correlation effects. DMC
    calculations on finite supercells are included to show
    convergence. The lines indicate a Birch--Murnaghan fit to the data.}
  \includegraphics[width=0.5\columnwidth]{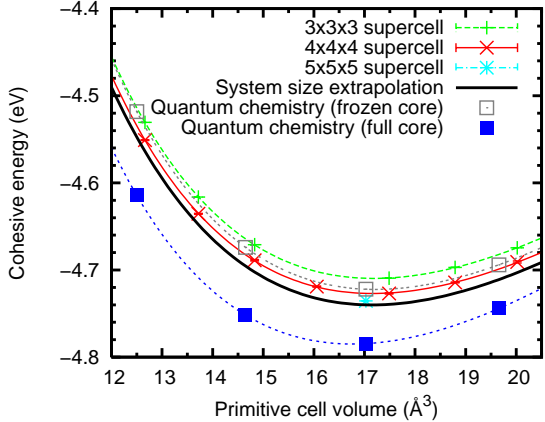}
  \label{graph:bulkpp}
\end{figure}

To study system-size errors, we calculated the DMC total
energy for several values of the atomic volume, using cubic supercells
containing $3 \times 3 \times 3$ and $4 \times 4 \times 4$ primitive
crystal cells (54 and 128 atoms). In addition, a DMC calculation on
the $5 \times 5 \times 5$ system (250 atoms) was performed at a single
atomic volume. The correction of 
Kwee \emph{et al.}~\cite{kwee_finite-size_2008}  was then applied
to each calculation using Eq.~(\ref{eq:kwee}), and two-point
extrapolation (Eq.~(\ref{eq:twopnt})) was used to reduce the remaining many-body
finite--size errors. To illustrate the effect of system-size
errors, we show in Fig.~\ref{graph:bulkpp} plots of $E_{\rm coh}$ as a
function of atomic volume from our DMC calculations on the
$3 \times 3 \times 3$ and $4 \times 4 \times 4$ supercells, as 
well as the results corrected for size errors, compared with our
earlier quantum-chemistry cohesive energies obtained with and without
core-valence correlation effects~\cite{nolan_calculation_2009}. 
In each case, we show also the
third-order Birch-Murnaghan fit to the results,
\begin{equation}
E(V)= E_0  -\frac{9 V_0 B_0}{16} \left(
(4-B'_{0})\frac{V_{0}^{3}}{V^{2}}-(14-3
B'_{0})\frac{V_{0}^{7/3}}{V^{4/3}}+(16-3B'_{0})\frac{V_{0}^{5/3}}{V^{2/3}} 
\right) \; ,
\end{equation}
where $E_0$, $V_0$ and $B_0$ represent the equilibrium
energy, volume and bulk modulus, respectively, and $B_0^\prime$ is the first
derivative of the bulk modulus.

The values of $E_{\rm coh}$, $a_0$ and $B_0$ obtained from the fit are
given in Table~\ref{table:bulkpp}. Several important points 
emerge from these results.
First, the system-size effects consist almost entirely of a
vertical shift, i.e. a constant energy offset, of the $E_{\rm coh} ( V )$
curves, so that they cause only small errors in $a_0$ and $B_0$.
For example going from the $3 \times 3 \times 3$ supercell to the
extrapolated system changes $a_0$ by only $\sim 0.1$~\%. Second,
comparing the raw values of $E_{\rm coh}$ from DMC on the $3 \times 3 \times 3$ and
$4 \times 4 \times 4$ supercells  to the fully-corrected and
extrapolated value suffers from substantial errors
of 133 and 52~meV, the KZK correction reduced these to 50 and 22~meV respectively. Third,
the DMC value of $E_{\rm coh}$, even after correction, still disagrees with
the quantum chemistry value of $E_{\rm coh}$ without core correlation energy
by $\sim 36$~meV. This last point indicates that the effect of 
the pseudopotential approximation
must be significant. In order to make further progress, all-electron
DMC is needed, and we report on this next.

\begin{table}[h]
\caption{Calculated bulk properties with both pseudopotential and
  all--electron DMC and 
  hierarchical quantum chemistry with and without core effects. The
  cohesive energy is calculated at 4.084\AA\  in all cases.}
\label{table:bulkpp}
\centering
\begin{tabular}{l r@{.}l r@{}l r@{.}l }
\hline\hline
 & \multicolumn{2}{c}{$a_0$ / \AA} & \multicolumn{2}{c}{$B_0$ / GPa}  & \multicolumn{2}{c}{$E_{\mathrm{coh}}$ / eV}  \\
\hline
DMC \supcell{3}\footnotemark[1]               &4&0965(2) &30&.5(1) & $-$4&6967(1)   \\
DMC \supcell{4}\footnote{Including the KZK correction}               &4&096(2) &31&.1(8) & $-$4&7249(1)   \\
DMC Extrap.                   &4&093(2)  &31&(1)   & $-$4&7466(3)   \\
DMC all--electron             &4&061(1)  &31&.8(4) &
$-$4&758(1)\footnote{This value is extrapolated to infinite--size,
  zero--timestep using six separate calculations with \supcell{3} and
  \supercell{4}s and timesteps of 0.004, 0.002 and 0.001 a.u.}    \\ 
Quantum Chemistry (no core)\cite{nolan_calculation_2009}    &4&099     &31&.9    & $-$4&7087      \\
Quantum Chemistry (with core)\cite{nolan_calculation_2009}  &4&062     &33&.1    & $-$4&7710      \\
Experiment\cite{nolan_calculation_2009}   &4&061(1)  &33&-38   & $-$4&778,$-$4.759\\
\hline\hline
\end{tabular}
\end{table}

\subsection{All-electron calculations}
\label{sec:all-electron}

In order to perform accurate all-electron DMC on the LiH crystal,
we have to address several technical challenges. First, as noted
in Sec.~\ref{sec:qmc}, the trial wavefunction must accurately
satisfy the Kato cusp condition at the nucleus, in order to
ensure stability of the walker population. Second, because of
the rapid variation of the orbitals near the nucleus, extremely
fine blip-grids are needed. Third, we expect to need much shorter
time-steps than for pseudopotential calculations. 

The technique
used to ensure that the cusp condition is satisfied is outlined
in Appendix~A. One symptom of the rapid variation of the orbitals
near the nucleus is the slow convergence of the DFT total energy with respect
to plane-wave cut-off in the {\sc quantum espresso} calculations
used to generate the orbitals. Given the high memory
requirements caused by the high cut-offs in the DMC calculations,
we took the orbitals to be sufficiently converged when they produced
stable walker populations. For our final all-electron DMC
calculations, we used orbitals generated using the LDA functional,
with a plane-wave cut-off of $6.8 \times 10^4$~eV. The associated
blip-grid had a spacing of half the natural grid dictated by the
plane-wave cut-off.

\begin{figure}[h]
  \caption{Total energy vs. time--step. This shows the convergence of
    the ae-DMC calculations with respect to time--step and is for a
    \supercell{3}. A linear extrapolation to zero timestep is included.}
  \includegraphics[width=0.5\columnwidth]{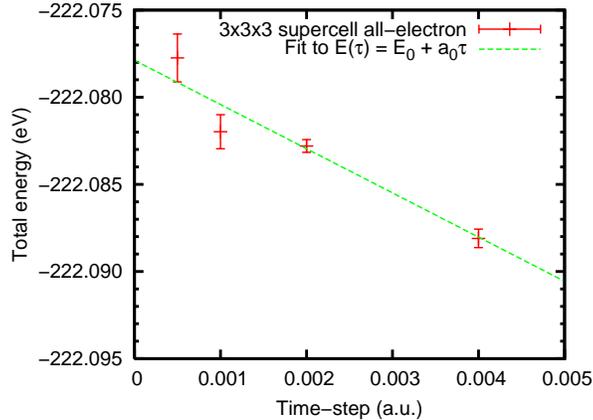}
  \label{graph:aetimestep}
\end{figure}

We made detailed tests on the time-step needed to ensure the accuracy
of the all-electron calculations. In Fig.~\ref{graph:aetimestep}, 
we show the results of tests
on the $3 \times 3 \times 3$ supercell, showing how the total energy
converges with respect to time step. The figure also shows a linear 
fit to the results, which is clearly adequate, as expected from earlier 
work.\cite{rothstein_time_1987} A time step
of $0.004$~au gives an error of only 10~meV/formula unit, 
which more than suffices to give accurate results for
the equilibrium $a_0$ and $B_0$.

In order to obtain the best possible results for the cohesive energy
as a function of volume $E_{\rm coh} ( V )$, we used the fact made
clear in Sec.~\ref{sec:pseudo} that results with 
the $3 \times 3 \times 3$ supercell
differ only by an almost constant energy offset from results converged
with respect to supercell size, this offset being in the region of 30~meV.
Our procedure in the all-electron DMC calculations was therefore to calculate
$E_{\rm coh} ( V )$ first with the $3 \times 3 \times 3$ supercell and
a time step of $0.004$~au. We then added a constant correction energy
to these results,  obtained from DMC calculations
with time steps of $0.001$, $0.002$ and $0.004$~au, all performed on
both the $3 \times 3 \times 3$ and $4 \times 4 \times 4$ supercells.
At each time step, the usual two-point 
extrapolation (Eq.~(\ref{eq:twopnt})) to
infinite supercell size was made, and a final linear time-step
extrapolation was then made.

\begin{figure}[h]
  \caption{Cohesive energy vs. primitive cell volume. Here the DMC
    energy differences have been calculated using a \supercell{3} and
    timestep of 0.004 a.u. and the point at 17.03\AA$^3$ calculated
    using the extrapolation procedure in the text.}
  \includegraphics[width=0.5\columnwidth]{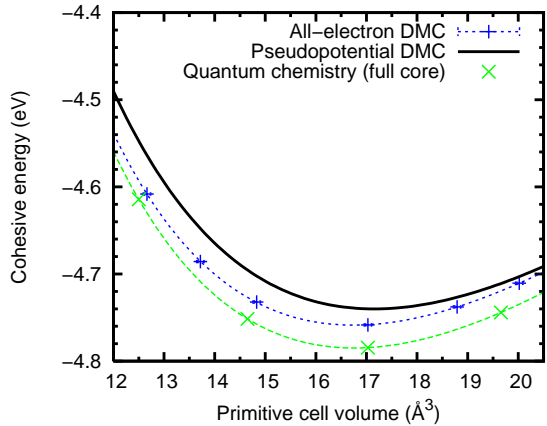}
  \label{graph:bulkae}
\end{figure}

The cohesive energy curve from all-electron DMC is 
compared in Fig.~\ref{graph:bulkae}
with our pseudopotential DMC curve and the results from quantum
chemistry. The resulting equilibrium values of $E_{\rm coh}$, $a_0$ and
$B_0$ are compared in Table~\ref{table:bulkpp}. We 
see that the all-electron DMC value
of $a_0$ agrees with the experimental and quantum chemistry values
to within $\sim 10^{-3}$~\AA\ ($0.03$~\%). Values of $B_0$ are much
more difficult to obtain accurately, but the all-electron DMC value agrees
with quantum chemistry to within $\sim 4$~\%, and both are reasonably
consistent with experimental values, which span a range of $\sim 15$~\%.
The all-electron DMC and quantum chemistry values of the equilibrium
$E_{\rm coh}$ differ by $13$~meV/formula unit. The
quantum chemistry value is believed to be somewhat more accurate than
this, so that some of this $13$~meV may be due to fixed-node error.\\

\section{Surface formation energy of LiH with QMC}
\label{sec:surf_qmc}
The methods of Secs.~\ref{sec:qmc} and \ref{sec:bulk_and_surface} have been used to calculate the formation
energy of the LiH~(001) surface, first with pseudopotentials, then
with all-electron DMC. All the calculations were done with the
lattice parameter $a_0 = 4.084$~\AA.

\subsection{Pseudopotential calculations}
\label{sec:surf_pseudo}

Exactly the same pseudopotential methods were used for the calculations
on slabs as were used in the bulk calculations of Sec.~\ref{sec:pseudo}, and the
trial orbitals were generated using DFT with the LDA functional, as before.
These orbitals were then re-expanded in B-splines using a spacing
corresponding to $\pi/2 k_{\rm max}$ where $k_{\rm max}$ is the
modulus of the largest plane-wave vector. For each surface supercell and each number $N$ of
ionic layers, the Jastrow factor of 1-, 2- and 3-body terms was
re-optimized using variance minimization.

To extract the values of $\sigma$ and $E_{\rm layer}$ for each chosen
surface unit cell, we performed calculations of the total slab
energy $E_{\rm slab} ( N )$ for numbers of ionic layers from 3 to 6
using a $4 \times 4$ surface unit cell (18 ions per layer in the
repeating supercell).  Single-particle size errors were corrected for
using Eq.~(\ref{eq:single}) with $a$ set equal to 1. Table
\ref{tab:surfconv} shows the convergence of $\sigma$ with respect to
the slabs used when fitting to Eq.~(\ref{eq:surf}).
 We have also performed calculations for
slabs 3 and 4 using a $3 \times 3$ surface unit
cells (18 ions per layer in the repeating supercell). Comparing
directly the $\sigma_{3,4}$ from the two different surface unit cells
differed by only 0.006(5) J\,m$^{-2}$ indicating the finite--size error. 
The resulting best value for $\sigma$ from the pseudopotential
calculations is 0.373~J\,m$^{-2}$. 

\begin{table}[h]
\caption{pp-DMC surface formation energy calculated using slabs of
  different thicknesses. Calculations performed on $4 \times 4$ surface
  unit cells.}
\centering
\begin{tabular}{l r@{}l}
\hline
\hline
Slabs used & \multicolumn{2}{c}{$\sigma~/ $ J\,m$^{-2}$}\\
\hline
3,4,5,6 & 0&.369(2)\\
4,5,6   & 0&.373(3)\\
5,6     & 0&.379(6)\\
\hline
\hline
\end{tabular}
\label{tab:surfconv}
\end{table}

\subsection{All-electron calculations}
\label{sec:surf_all-electron}

The all-electron DMC techniques used for the slab calculations
were essentially the same as those used for the 
bulk (Sec.~\ref{sec:all-electron}).
However, the memory requirements for the B-spline coefficients
were so much greater than for the bulk that we had to reduce the
plane-wave cut-off used to generate the orbitals from $6.8 \times 10^4$
to $3.4 \times 10^4$~eV. This primarily made the DMC runs more
susceptible to population control issues and resulted in a higher
statistical error on the final values compared to the pseudopotential work.
For the same reason, were able to perform all-electron calculations
only for the $3 \times 3$ surface unit cell, and the largest number
of ionic layers that we could handle was $N = 5$. We know from the
pseudopotential calculations that the finite size errors are under
control using $3 \times 3$ surface unit cells assuming the LDA
correction is used. The introduction of the
tightly bound core electrons is not expected to 
increase the finite-size errors.

A further technical issue in the all-electron slab calculations was
concerned with the optimization of the Jastrow factor. In order to
obtain wavefunctions that produced stable DMC runs we found it
necessary to optimize the Jastrow factor using the energy minimisation
scheme within VMC. This tended to increase the variance of
the local energy of the trial wavefunction slightly with respect to
variance minimisation. However it did reduce the number of population
explosions during the DMC runs.

The timestep adopted ($0.004$~a.u.) was the same as 
used in the all-electron bulk
work, since the tests done there indicated that this is sufficient. 
Our final all-electron DMC result for 
the surface formation energy is $\sigma=0.44(1)~\rm J\,m^{-2}$.
We note that the explicit inclusion of Li core states increases $\sigma$ by
$\sim 0.07$~J\,m$^{-2}$, which is a significant effect 
at the level of accuracy sought in this work.\\

\section{Hierarchical quantum chemistry for surface energy}

 \label{sec:surf_hierarch}

The formation energy of the of LiH~(001) surface was also
computed using quantum chemistry techniques.  The Hartree-Fock
component $\sigma^\mathrm{HF}$ was determined from slab calculations,
see Eq.~(\ref{eq:surf}), using the CRYSTAL and VASP
codes.  The effect of electron correlation was accounted for using the
hierarchical method as described in Section~\ref{sec:techqc}. 

Both CRYSTAL and VASP employ periodic boundary conditions, so
that the calculations are performed on an infinite array of slabs,
with a vacuum gap separating successive slabs. 
The vacuum gap was chosen to be 26~\AA, large enough to ensure that there is
no interaction between neighbouring slabs.
Careful attention was also paid to convergence with respect to
$k$-point sampling and basis-set completeness.
A previous high accuracy Hartree-Fock study of bulk LiH was
performed using CRYSTAL by Paier et al\cite{paier_accurate_2009}.  
The basis set
described in that work was used for the present calculations.  In order to
ensure basis set completeness, layers of ``ghost'' atoms were added
above and below each surface.  The ``ghost'' atoms were basis functions
centred on the sites of atoms in the next layer but without the nuclei
or electrons.  The convergence of $\sigma^\mathrm{HF}$ with respect to
these ``ghost" atoms was tested using slabs of four and five layers,
see Table \ref{tab:ghost}.  The introduction of the ``ghost" atoms has a
significant effect on $\sigma^\mathrm{HF}$ and two layers are
necessary to achieve basis set completeness; this number of layers was
used in all our calculations.

\begin{table}[h]
\begin{center}
\caption{Convergence of $\sigma_\mathrm{HF}$ using CRYSTAL with
  respect to the number of layers of ``ghost'' atoms above and below
  the surface. Based on two-point extrapolations from slabs of 4-5
  layers.  All energies are quoted 
  in J\,m$^{-2}$.} 
\begin{tabular}{lr}
\hline\hline
Ghost layers & $\sigma_\mathrm{HF}$  \\
\hline
0 & 0.43835 \\
1 & 0.19886 \\
2 & 0.19849 \\
3 & 0.19854 \\
\hline\hline
\end{tabular}
\label{tab:ghost} 
\end{center}
\end{table}

Calculations on slabs of two to eight layers were performed using both
periodic codes, and the method outlined in Sec.~\ref{sec:surf_pseudo}
was used to extract values of $\sigma^{\rm HF}$. The resulting values
are shown in Table~\ref{hf_tab}.  We note that the VASP value is
slightly lower than the CRYSTAL value. Since CRYSTAL provides
a direct all-electron calculation, and we have established that 
the CRYSTAL result is
converged with respect to basis set, we suggest that the VASP value
may be a slight underestimate. This may be due to
the PAW potentials used: the standard PBE potentials were used, and
while harder potentials are available for H, it was not possible to
reach convergence with these potentials.  Previous studies of bulk LiH
with VASP have reported small discrepancies in the Hartree-Fock 
result~\cite{marsman_second-order_2009}.  The CRYSTAL 
results converge with respect
to slab thickness to give a value of 0.198(1)~J\,m$^{-2}$. 

\begin{table}[h]
\begin{center}
\caption{Hartree-Fock approximation to surface formation energy for
  LiH, $a=4.084$~\AA, using CRYSTAL and VASP.  All energies are quoted
  in J\,m$^{-2}$.} 
\begin{tabular}{lcl}
\hline\hline
Slabs used & CRYSTAL & VASP \\
\hline
2-8 & 0.20005 & 0.19363\\
3-8 & 0.19883 & 0.19114\\
4-8 & 0.19825 & 0.19001\\
5-8 & 0.19819 & 0.18944\\
6-8 & 0.19836 & 0.18926\\
7-8 & 0.19703 & 0.18864\\
\hline\hline
\end{tabular}
\end{center}
\label{hf_tab}
\end{table}

\begin{figure}[h]
  \centering
  \caption{The convergence of $E^{011}$ (red circles) and $E^{111}$
    (blue squares) with respect to maximum cluster size $N$ using
    MP2/cc-pVTZ for LiH $a = 4.084$~\AA} 
  \includegraphics[width=0.5\columnwidth]{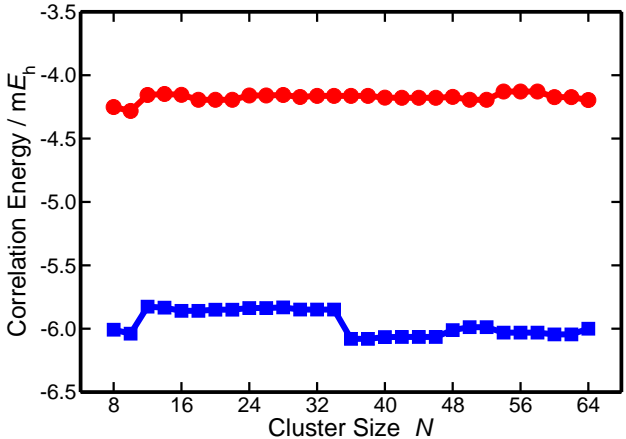}
  \label{graph:hier}
\end{figure}

The correlation component of the surface formation energy was
calculated using the hierarchical method.  The convergence of the
hierarchical coefficients is shown in Fig.~\ref{graph:hier} and
using the methods described in Ref.~\onlinecite{nolan_calculation_2009} 
the values can
be converged to within a few tenths of mE$_\mathrm{h}$.
In brief, a reference calculation was performed using $N=64$ and frozen-core MP2 theory in the cc-pVTZ basis set. 
Corrections for core correlation ($\delta$core), basis-set incompleteness ($\delta$basis) and higher-level correlation treatments ($\delta$CCSD(T), $\delta$CCSDT, $\delta$CCSDT(Q)) and the diagonal Born-Oppenheimer correction were also computed using smaller basis sets and smaller values of $N$. The use of smaller values of $N$ for small corrections was validated in earlier 
work~\cite{nolan_calculation_2009}.
Complete details of the hierarchical results are given in 
Table~\ref{hier_tab}.
An error of 0.2~mE$_\mathrm{h}$ in the hierarchical coefficients 
corresponds to an
error of 0.005~J\,m$^{-2}$ in the surface formation energy. 
To facilitate comparison with both DMC results, $\sigma^\mathrm{corr}$
has been calculated with and without correlating the core electrons.

\begin{table}[h]
\begin{center}
\caption{
Correlation contributions to the surface formation energy of LiH at $a=4.084$~\AA\ and total calculated surface formation energies with and without core correlation. In each row, the $N$ value specifies the maximum number of ions in the hierarchical calculation. Correlation-consistent basis sets have been used throughout, and cc-p(C)V$X$Z is abbreviated (C)V$X$Z. 
The CCSDT and CCSDT(Q) calculations were performed using MRCC  \cite{kallay,mrcc} as a module in Molpro and the diagonal Born-Oppenheimer correction (DBOC) calculations were performed using PSI3 \cite{psi3}. All other calculations were performed using Molpro \cite{molpro}. 
} 
\begin{tabular}{lccrlr}
\hline\hline
reference  & $E^{011}$/m$E_{\rm h}$ & $E^{111}$/m$E_{\rm h}$ & $\sigma^\mathrm{corr}$/J\,m$^{-2}$ & details\\
\hline
 & $-4.196$ & $-6.000$ & +0.1886 &MP2/VTZ $N=64$\\
$\delta$core  & $-0.840$ & $-1.147$ & +0.0321 & MP2/CVTZ $-$ MP2/VTZ $N=16$\\
$\delta$basis  & +0.268 & +0.268 & $+0.0000$ & MP2/V[T,Q]Z $N=36$\\
  & +0.162 & +0.158 & $+0.0004$ & MP2/V[Q,5]Z $N=16$\\
$\delta$CCSD(T)  & +0.625 & +0.538 & +0.0091 & CCSD(T)/VTZ $-$ MP2/VTZ $N=16$\\
$\delta$CCSDT  & $-0.149$ & $-0.209$ & +0.0063 & CCSDT/VDZ $-$ CCSD(T)/VDZ $N=8$\\
$\delta$CCSDT(Q)  & $-0.015$ & $-0.023$ & +0.0008 & CCSDT(Q)/VDZ $-$ CCSDT/VDZ $N=8$\\
DBOC  & +0.041 & +0.055 & $-0.0015$ & HF/VTZ $N=8$\\
\hline
$\sigma^\mathrm{corr}_\mathrm{frozen~core}$ & & & 0.2037 & Sum terms above except $\delta{}$core \\
$\sigma^\mathrm{corr}_\mathrm{total}$ & & & 0.2358 & Sum all terms above \\
\hline
$\sigma^\mathrm{HF}$ & & & $0.198\phantom0$ &  \\
$\sigma^\mathrm{static}_\mathrm{frozen~core}$ & & & $0.402\phantom0$ &  \\
$\sigma^\mathrm{static}_\mathrm{total}$ & & & $0.434\phantom0$ &  \\
\hline\hline
\end{tabular}
\end{center}
\label{hier_tab}
\end{table}

\section{Discussion} 
\label{sec:discussion}

We summarize in Table~\ref{table:surfacedmc} our QMC and hierarchical
quantum chemistry results for the formation energy $\sigma$ of
the LiH~(001) surface, and we compare with the predictions of DFT
using the LDA, PBE and rPBE functionals, these DFT results being
taken from our earlier work~\cite{binnie_benchmarking_2009}.
The very close agreement between the DMC and QC results for $\sigma$
confirms that both approaches give high accuracy, and shows
that the results can be used as benchmarks for assessing DFT
approximations. The DFT values of $\sigma$ span a remarkably
wide range, with LDA overestimating it by $7$~\%, and PBE and
rPBE underestimating it by $23$~\% and nearly $40$~\% respectively.
It is interesting to note that the Hartree-Fock value $\sigma^{\rm HF}$
of $0.20$~J~m$^{-2}$ (Table~\ref{hf_tab}) accounts for less than
half the full value of $\sigma$, so that the importance of an accurate
treatment of correlation is clear. It is also noteworthy that
valence-core correlation gives a surprisingly significant
contribution of $\sim 10$~\% to $\sigma$. We noted in the Introduction
the scattered evidence that LDA tends to give better values of $\sigma$
than GGA approximations, and the present work shows that this is the
case for LiH.

\begin{table}[h]
\caption{
Calculated surface formation energy with both pseudopotential and
  all--electron DMC and 
  hierarchical quantum chemistry with and without core effects. The
  calculations are performed at 4.084\AA\  in both cases. DFT data
  from previous work is included. The DFT values are for the lattice
  parameters optimized with the given 
  functionals.\cite{binnie_benchmarking_2009}}
\label{table:surfacedmc}
\centering
\begin{tabular}{l r@{.}l}
\hline\hline
method & \multicolumn{2}{c}{$\sigma$ / J\,m$^{-2}$}\\
\hline
DMC pseudopotential           &0&373(3)\\
DMC all--electron             &0&44(1) \\
Quantum Chemistry (froz. core)   &0&402  \\
Quantum Chemistry (with core) &0&434  \\
DFT LDA                       &0&466  \\
DFT PBE                       &0&337  \\
DFT rPBE                      &0&272  \\
\hline\hline
\end{tabular}
\end{table}

An important part of the evidence that our DMC and hierarchical QC
calculations give results of benchmark quality for $\sigma$
is the very high accuracy of the two completely
independent approaches for the energetics of the LiH crystal; for
hierarchical quantum chemistry, this was already shown in detail
in Ref.~\onlinecite{nolan_calculation_2009}, and a substantial part of the present paper has been devoted to
showing the same thing for QMC. In both approaches, the calculated
cohesive energy is correct to $\sim 15$~meV/formula unit, and the equilibrium
lattice parameter to within better than $0.1$~\%. It is clear from both
sets of calculations that an adequate treatment of core-valence
correlation must be included. The reliability of the calculated values
for $\sigma$ is confirmed by the very close agreement (within
$\sim 0.01$~J\,m$^{-2}$) between the values given by the two approaches. Here
too, core-valence correlation is important, giving a contribution
of $\sim 0.03$~J\,m$^{-2}$ to $\sigma$.

In both theoretical approaches, a key technical issue is system size effects.
In QMC, the calculations are done directly in periodic boundary conditions,
but large repeated systems are needed because we rely on $\Gamma$-point
sampling. For the bulk crystal, we have shown that significant
corrections need to be made for size errors, but that these are successful
in reducing the errors in the cohesive energy to $\sim 15$~meV/formula unit.
For the DMC slab calculations of $\sigma$, we have presented evidence
that the calculated $\sigma$ is very well converged (to within
0.006~J\,m$^{-2}$) with respect to both slab thickness and size of the
surface repeating unit. In the hierarchical approach, the Hartree-Fock
part of $\sigma$ is calculated in periodic boundary conditions, and we have
shown that size errors in the slab calculations can be made negligible.
Size errors in the correlation residual contributions are well 
controlled in the hierarchical scheme, and, as can 
be seen from Fig.~\ref{graph:hier}, are of the order of 
0.1 ${\rm m}E_{\rm h}$ per ion.

It would now be timely to extend the present calculations to other
materials. In fact, we have reported QMC calculations of
$\sigma$ for MgO~(001) several years ago~\cite{alf_energetics_2006}, though it might be worth repeating
the calculations with the improved pseudopotentials now available.
Our hierarchical quantum chemistry scheme can be applied without change to
MgO and other materials having the rock-salt structure, and we hope
to report both QMC and hierarchical calculations on LiF in the near
future. However, it is important to emphasise that the hierarchical
scheme also works well for materials having other crystal structures, so that
there is now rather wide scope for using it, with or without QMC,
for calculations of $\sigma$. We remark that for some materials
it will be essential to include the effects of surface relaxation.
This is not a significant issue for most rock-salt materials, and we
have shown\cite{binnie_benchmarking_2009} that for LiH relaxation reduces $\sigma$ by only
$\sim 0.01$~J\,m$^{-2}$. But in other cases (corundum is a famous
example), relaxation makes a large difference to $\sigma$, and would
probably need to be estimated from DFT calculations.

To conclude, we have shown that: (a) quantum Monte Carlo calculations
give extremely accurate results for the energetics of the LiH crystal,
particular when Li core electrons are explicitly included, and there
is excellent agreement with results from the hierarchical quantum
chemistry scheme; (b) these two independent techniques give almost
identical benchmark results for the formation energy of the LiH~(001)
surface; (c) the benchmark value of $\sigma$ lies between DFT predictions
from the LDA and GGA approximations, the LDA value being somewhat better
than GGA.\\

\section*{Acknowledgements}
This research used resources of the Oak Ridge Leadership Facility at
the Oak Ridge National Laboratory, which is supported by the Office of
Science of the U.S. Department of Energy under Contract
No. DE-AC05-00OR22725. The UKCP consortium provided time on HECToR,
the UK's national high-performance computing service, which is
provided by UoE HPCx Ltd at the University of Edinburgh, Cray Inc and
NAG Ltd, and funded by the Office of Science and Technology through
EPSRC's High End Computing Programme. S.J.B. acknowledges an EPSRC
studentship. S.J.N. acknowledges funding from the School of Chemistry
at the University of Bristol. At the time most of this research was
performed, F.R.M. was funded by the Royal Society. The authors are
immensely grateful to Kirk Peterson for providing by return the
core-valence basis sets for lithium that were used in this
research. The work of D.A. was conducted as part of a EURYI scheme see 
www.esf.org/euryi.\\

\appendix
\section{Generalised cusp correction}
\label{app:cusp}
A scheme for modifying real orbitals expanded in a Gaussian basis set so that
they satisfy the Kato cusp conditions\cite{kato_eigenfunctions_1957,pack_cusp_1966} is described in Ref.\
\onlinecite{ma_scheme_2005}.  In the present work we make use of an extension of
this scheme which allows the Kato cusp conditions to be imposed on complex
orbitals expanded in any smooth basis set.

For simplicity, we restrict our attention to the case of an all-electron
nucleus of charge $Z$ at the origin in the following discussion.  In the
scheme of Ref.\ \onlinecite{ma_scheme_2005}, the s-type Gaussian basis functions are
replaced by radial functions in the vicinity of the nucleus.  These functions
impose the cusp conditions and make the single-particle local energy resemble
an ``ideal'' curve that was found empirically to be satisfied by a wide range
of Hartree-Fock atomic orbitals.  In the scheme used in this work, instead of
\textit{replacing} part of the orbital, we \textit{add} a spherically
symmetric function of constant phase to the orbital.  The function added to
uncorrected orbital $\psi({\bf r})$ is
\begin{equation} \Delta \psi(r) = \exp(i\theta_0) \left[ \tilde{\phi}(r) -
    \phi(r) \right] \Theta(r_c-r),
\end{equation} where $\Theta$ is the Heaviside function, $r_c$ is a cutoff
length, $\theta_0=\arg[\psi({\bf 0})]$,
\begin{equation} \phi(r)={\rm Re} \left[ \frac{\exp(-i\theta_0)}{4\pi}
    \int_{\rm sphere} \psi({\bf r}) \, d\Omega \right], \end{equation} and
\begin{equation} \tilde{\phi}(r)=C+\exp(\alpha_0+\alpha_1 r+\alpha_2 r^2
  +\alpha_3 r^3 + 
  \alpha_4 r^4), \end{equation} where $C$ is a real constant and the
$\{\alpha\}$ are real constants to be determined.  In practice $\phi(r)$ is
evaluated by cubic spline interpolation; the spherical averaging of the
uncorrected orbital is performed on a radial grid at the outset of the
calculation.  $C$ is chosen so that $\phi(r)-C$ is positive everywhere within
the Bohr radius of the nucleus.

The uncorrected orbital may be written as
\begin{equation} \psi({\bf r})=\exp(i\theta_0) \phi(r)+\eta({\bf r}),
\end{equation}
where $\eta({\bf r})$ consists of the $l>0$ spherical harmonic components of
$\psi({\bf r})$, together with the phase-dependence of the $l=0$ component.
Note that $\exp(i\theta_0)\phi(0)=\psi({\bf 0})$, and hence $\eta({\bf 0})=0$.
We may now apply the scheme of Ref.\ \onlinecite{ma_scheme_2005} to determine $\{
\alpha \}$ and $r_c$, with $\exp(i\theta_0) \phi$ and $\exp(i\theta_0)
\tilde{\phi}$ playing the roles of the uncorrected and corrected s-type
Gaussian functions centered on the nucleus at the origin.  The constant phase
$\exp(i\theta_0)$ cancels out of the equations that determine the $\{\alpha\}$
[Eqs.\ (9)--(13) in Ref.\ \onlinecite{ma_scheme_2005}], so the determination of the
$\{\alpha\}$ and $r_c$ is exactly as described in Ref.\ \onlinecite{ma_scheme_2005},
except that we do not need to modify $Z$ when more than one nucleus is
present, because $\eta({\bf 0})=0$.

Suppose the orbital is of Bloch form $\psi({\bf r})=u_{\bf k}({\bf
r})\exp(i{\bf k}\cdot {\bf r})$, where $u_{\bf k}$ has the periodicity of the
primitive cell.  Let $\{{\bf R}_p\}$ be the set of primitive-cell lattice
points.  The orbital may be corrected at each all-electron nucleus in the
primitive cell at ${\bf R}_p={\bf 0}$ using the scheme described above.  The
phase of the orbital (and hence cusp-correction function) at the corresponding
nucleus in the primitive cell at ${\bf R}_p \neq {\bf 0}$ is $\exp(i{\bf k}
\cdot {\bf R}_p)$ times that for the primitive cell at ${\bf R}_p={\bf 0}$;
the cusp-correction function is otherwise identical.  The corrected orbital is
of Bloch form.

\bibliographystyle{aip}
\bibliography{citations}

\end{document}